\documentclass[12pt]{article}
\usepackage{epsfig}
\def\e{\begin{equation}}
\def\f{\end{equation}}
\def\=#1{\overline{\overline{#1}}}
\def\_#1{{\bf #1}}
\def\o{\omega}
\def\E{\epsilon}
\def\M{\mu}
\def\.{\cdot}
\def\x{\times}

\def\l#1{\label{eq:#1}}
\def\r#1{(\ref{eq:#1})}
\def\d{\nabla}

\def\ds{\displaystyle}

\def\l#1{\label{eq:#1}}
\def\r#1{(\ref{eq:#1})}

\renewcommand{\Re}{\mathop{\rm Re}\nolimits}

\title{The planar perfect lens: physical requirements and possible alternative
realizations} \author{Stanislav Maslovski, Sergei Tretyakov\\
Radio Laboratory, SMARAD, Helsinki University of Technology\\
P.O. Box 3000, FIN-02015 HUT, Finland\\[4mm]
E-mails: stanislav.maslovski@hut.fi, sergei.tretyakov@hut.fi}
\date{\today}

\begin{document}
\maketitle

\begin{abstract} Several alternative possibilities of how to
create an  electromagnetic device being able to reconstruct
near-field distribution of a source with sub-wavelength resolution
(so-called {\em perfect lens}) are considered. It is shown that there is
a variety of such means not involving double-negative (left-handed, or {\it Veselago})
materials or periodical backward-wave structures.
It is demonstrated that devices working in a similar manner can be constructed
using planar grids or material sheets imposing necessary boundary conditions
at two parallel planes in air.
\end{abstract}

\section{Introduction}

It is known that using materials with
simultaneously  negative permittivity and permeability (at a given
frequency) focusing of divergent homocentric electromagnetic beams
by ``planar lenses'' becomes possible because of the negative
refraction \cite{Veselago}. As it was found in
\cite{Pendry}, such lenses are also able to ``amplify'' evanescent
fields carrying information about sub-wavelength details of a
source near field. There were also a few experimental papers \cite{Smith1,Smith2}
demonstrating negative refraction effects in microwave composite
materials. These facts received a lot of attention (and criticism)
in the recent literature (see, e.g.,  \cite{Valanju,Garcia}).
The ``amplification'' of evanescent waves in a Veselago's slab
lens is a phenomenon that easily contradicts with intuition and common sense,
especially for those who are used to associate  the word
``amplification'' with an active device that amplifies the signal power.
As it was recently shown in \cite{comment,my}, this
``amplification'' is simply a resonant excitation of waveguide
modes of a slab waveguide filled by the Veselago medium.
Misunderstanding of the nature of this phenomenon comes, in our opinion,
from a not quite appropriate and clear terminology used
in the first papers on this subject. The terms ``resonance'' or ``resonant
growth'' are probably more appropriate for this phenomenon.

In this paper, we will not discuss  problems and difficulties in
theoretical interpretation and practical realization of backward-wave materials
and the perfect lens based on these media. Only briefly, we note that from our point of view,
the physical existence of  the negative
refraction phenomenon as well as the evanescent field
``amplification'' phenomenon is beyond any theoretical doubt today. Still,
there maybe some questions regarding the accuracy of the experimental
approach used in \cite{Smith2} and the degree of the performance
deterioration because of inevitable losses and the finite size of the
lens,  but at least theoretically the
problem may be considered now as well-understood.
Indeed, the causality question mentioned in \cite{Valanju} has
been solved in \cite{Schurig}. The negative influence of the
medium dispersion and losses discussed in
\cite{Garcia} can, in principle, be overcome by using
metamaterials involving active devices \cite{TretyakovMOTL}.

However, these difficulties call for a study of alternative realizations of
a device that can restore near fields.
Thus, despite the fact that there is now a direct and in principle
well-understood way to realizing new lenses by means of backward-wave materials,
we will make a step aside in this paper. Starting from the analysis
of an ideal Veselago slab lens, we will
formulate a couple of equivalent problems and show that the
special electromagnetic property of the material filling,
namely, existence of backward waves, is not, in fact, crucial for
a planar device operating as a lens or, moreover, as a perfect lens.
What makes a Veselago lens to behave as a perfect lens, are the properties of
the two slab interfaces.
We will show that devices working in a similar manner can be constructed
using planar grids or material sheets imposing necessary boundary conditions
at two parallel planes in air.

\section{The ideal Veselago lens and an equivalent problem}

Let us start from considering an ideal Veselago slab lens
operation.  Its well-known structure is depicted in Figure~\ref{fig1}.
We work in the frequency domain, and the time dependence is of the form
$e^{+j\omega t}$.
\begin{figure}[h] \begin{center} \begin{tabular}{c|c|c}
region 1 & region 2 & region 3\\
 & & \\
$\begin{array}{l}
\d\x\_E_1 = -j\o\M_0\_H_1\\
\d\x\_H_1 =  j\o\E_0\_E_1\\
\end{array}$
&
$\begin{array}{l}
\d\x\_E_2 =  j\o\M_0\_H_2\\
\d\x\_H_2 = -j\o\E_0\_E_2\\
\end{array}$
&
$\begin{array}{l}
\d\x\_E_3 = -j\o\M_0\_H_3\\
\d\x\_H_3 =  j\o\E_0\_E_3\\
\end{array}$\\
 & & \\
air & Veselago medium & air\\
\end{tabular}
\end{center}
\caption{Veselago slab lens: a planar slab of a backward-wave material
with the medium parameters $\epsilon=-\epsilon_0$ and $\mu=-\mu_0$ in free space.}
\label{fig1}
\end{figure}
We suppose that the lens is positioned in space with the relative
permittivity and permeability equal to $1$. The corresponding relative
parameters of the slab material both equal to $-1$ at the working frequency.
The boundary conditions at
the lens interfaces are the usual Maxwellian boundary conditions (the tangential
components of the fields are continuous across the interfaces). The corresponding
field equations are also shown in Figure~\ref{fig1} for all three regions.

It  is easy to notice that the equations in region 2
differ from that in regions 1 and 3 only by complex conjugation. Substitution
\e \_E_{\rm
(old)}, \_H_{\rm (old)} \Rightarrow \_E_{\rm (new)}^*, \_H_{\rm (new)}^*
\l{subst} \f
(here and thereafter $^*$ denotes the complex conjugation
operation) into the field equations in region 2 results in a new but
equivalent problem written for new field vectors, as it is shown in
Figure~\ref{fig2}.
\begin{figure}[h] \begin{center}
\begin{tabular}{c|c|c}
region 1 & region 2 & region 3\\
$\begin{array}{rr}
 & \_E_{\rm t_1}\\
\E_{\rm r_1}=1 & \\
\M_{\rm r_1}=1 & \\
 & \_H_{\rm t_1}\\
\end{array}$
&
$\begin{array}{lrr}
\_E_{\rm t_2}^* & & \_E_{\rm t_2}^*\\
 & \E_{\rm r_2}=1 & \\
 & \M_{\rm r_2}=1 & \\
\_H_{\rm t_2}^* & & \_H_{\rm t_2}^*\\
\end{array}$
&
$\begin{array}{lr}
\_E_{\rm t_3} & \\
 & \E_{\rm r_3}=1\\
 & \M_{\rm r_3}=1\\
\_H_{\rm t_3} & \\
\end{array}$\\
air & air & air\\
\end{tabular}
\end{center}
\caption{Two conjugating planes in free space. This system is equivalent to that
shown in Figure~\ref{fig1}.}
\label{fig2}
\end{figure}

In the new formulation   the field equations are the same in all three regions,
and they
are simply the Maxwell equations in free space:
\e \d\x\_E =
-j\o\M_0\_H,\qquad \d\x\_H = j\o\E_0\_E \f
The boundary conditions on the two interfaces, however, are
no more the standard continuity conditions, but they involve
complex conjugation:
\e \_E_{\rm
t_{(1,3)}}=\_E_{\rm t_{(2)}}^*, \qquad \_H_{\rm t_{(1,3)}}=\_H_{\rm t_{(2)}}^*
\l{BC} \f
Let us discuss the physics of
these boundary conditions involving complex conjugation a bit
later. At this stage we see that an ideal Veselago slab refraction
problem is mathematically equivalent to the refraction at two
conjugating planes in free space. Hence, for the system of two
such planes in free space the  field solutions are the same as for a
Veselago material slab:  propagating plane waves are refracted negatively, and
the evanescent modes are ``amplified'', which are, obviously, the conditions for a
perfect lens.

\begin{figure}[h] \begin{center} \epsfig{file=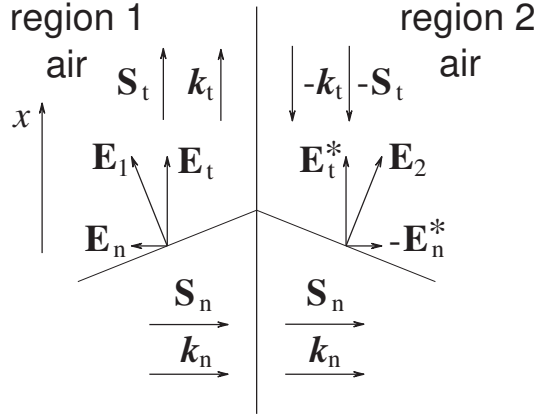,width=7cm}
\end{center} \caption{A single conjugating interface. A  TM-polarized
plane wave is incident from the left.} \label{fig3}
\end{figure}

To  understand how these conjugating planes operate
consider a plane wave incidence problem for a single conjugating plane. The
problem geometry for the TM incidence is shown in Figure~\ref{fig3}.
The incident wave comes from
the left. The magnetic field is orthogonal to the picture plane and is
not shown.
Because of the specific nature of the boundary
conditions \r{BC}, the solution for the transmitted wave must
depend on the tangential coordinate $x$ as $e^{+jk_{\rm t}x}$
($k_{\rm t}$ denotes the tangential component of the wave vector),
if the incident wave phasor is $e^{-jk_{\rm t}x}$, otherwise one
cannot satisfy the boundary conditions. We can say that $k_{\rm t}$
changes sign when a plane wave passes across the interface.
In this process, the transmitted field  energy in the second
region must propagate to the right. This determines the direction
of the Poynting vector normal component $\_S_{\rm n}$. The second
region is now an air region that means the normal component of the
wave vector in the second region is along the same direction as
$\_S_{\rm n}$. A similar relation holds for $\_k_{\rm t}$, $\_S_{\rm
t}$ pair (see the picture). Clearly, negative refraction
takes place.

It is possible to write simple equations for the complex field
amplitudes of the reflected and transmitted fields. If $A$, $B$, and
$C$ denote the amplitudes of the incident, transmitted, and
reflected wave electric field tangential components, respectively, then for the
considered TM case one can write:
\e \begin{array}{ccc}
A + C &=& B^* \\
(A - C)/\eta &=& B^*/\eta^*\\
\end{array} \l{abc} \f
Here $\eta=\eta(k_{\rm t})$ is the wave
impedance connecting tangential components of electric and
magnetic fields of a wave. It is obviously a function of $k_{\rm
t}$. For the propagating modes $\eta$ is purely real, for the
evanescent ones it is purely imaginary.
The solution of \r{abc} is
\e
C = {\ds 1-\eta/\eta^*\over \ds 1+\eta/\eta^*}A, \qquad B = {\ds 2A^*\over \ds 1+\eta^*/\eta}
\l{abcsol}
\f

It is seen that for propagating modes (real wave impedance) the
interface is perfectly matched: reflected field amplitude $C = 0$,
and the transmitted field amplitude $B = A^*$. For evanescent modes (imaginary wave
impedance) the denominator of \r{abcsol} is zero, and a surface
wave (surface polariton) resonance occurs: $B, C\rightarrow \infty$.
As is known \cite{comment,TretyakovMOTL},
an interface between free space and a backward-wave material
with $\epsilon=-\epsilon_0$ and $\mu=-\mu_0$ has similar properties.
Another interesting feature
of the obtained result is that it is impossible to introduce
the usual transmission coefficient because the transmitted field
is proportional to $A^*$, but not directly to $A$. This is
because the boundary conditions \r{BC} are no more linear in the sense of
multiplication by a complex number.

An important conclusion from this analysis is that all the
phenomena necessary for perfect reconstruction of the entire
wave spectrum take place {\it at the two interfaces}
of the Veselago slab. If one can realize a sheet such that traveling waves
refract negatively when crossing this sheet, a system of two such sheets
{\it in free space}
will focus propagating modes of a source just like a Veselago slab.
If this sheet also supports surface waves for any $k_t > k_0=\o\sqrt{\E_0 \M_0}$,
then two such sheets in free space will reconstruct the entire evanescent spectrum
as well.

\section{Physical restrictions}

Let us now discuss the physical meaning of the boundary conditions \r{BC}
and the restrictions on their physical realization. Complex conjugation in the
frequency domain corresponds to time reversal in the time domain.
If the boundary conditions \r{BC} were true for every spectral
frequency component in the range $(-\infty, +\infty)$, then this
condition would obviously violate the causality principle.
Speaking in simple words, one could in this case say that
transformation \r{subst} forces the time to go in the opposite
directions inside sub-regions of a single physical system. But if
we are interested only in steady-state operations at a given
frequency, then the complex conjugation becomes possible, at least
using nonlinear or active devices, e.g., mixers.
Conceptually, if a thin sheet of a nonlinear material is
illuminated by a signal plane wave with the harmonic time
dependence $\cos(\omega_0 t+\phi)$ and a
reference plane wave  $\cos(2\omega_0 t)$, among the
output harmonics there is a plane wave $\cos(\omega_0 t-\phi)$,
which corresponds to the complex conjugation of the original
field.

Another important point is the following. Applying transformation
\r{subst} and concluding that the middle region has became a true
free-space region we have silently assumed that {\it all}
electromagnetic quantities and relations have been kept in their
original free-space form after such a transformation. Let us
consider an illustrative example. A simpler idea of how to
transform the field equations of an ideal Veselago slab to
free-space equations is changing the sign of the electric (or magnetic)
field in the Maxwell equations. The (sourceless) field equations take the necessary
form after such transformation, but the problem is that the
Poynting vector $\_S=\Re(\_E\x\_H^*)$ in the old variables becomes
$\_S=-\Re(\_E\x\_H^*)$ in the new ones. From the other hand, it is
easy to see that substitution \r{subst} preserves the Poynting
vector expression in the original form. The physical reason of
this is, of course, the time reversal invariance of a reciprocal
electromagnetic system.

\section{Possible realizations not involving complex conjugation of the fields}

In  the previous section we have considered a potential device
that is able to ideally imitate the operation of a Veselago slab lens. Now
we will concentrate on other possibilities providing additional
freedom in realization of sub-wavelength resolution lenses without
involving the complex conjugation operator.
In this section we will make use of  a powerful
synthesis method based on so-called transmission matrices, known in the
microwave circuit theory.
These matrices connect the complex amplitudes of waves traveling in the
opposite directions and measured at two reference planes:
\e \left(\begin{array}{c}
E_2^{-}\\
E_2^{+}\\
\end{array}\right)
=\left(\begin{array}{cc}
t_{11} & t_{12}\\
t_{21} & t_{22}\\
\end{array}\right)
\.
\left(\begin{array}{c}
E_1^{-}\\
E_1^{+}\\
\end{array}\right) \f
Here, $E_{1}^{\pm}$ and $E_{2}^{\pm}$
denote the tangential components of the electric field complex amplitudes
of  waves at
the first (input) and the second (output) interfaces of a device,
respectively (we restrict ourselves by plane structures and plane
waves). The signs $^\pm$ correspond to the signs in the propagator
exponents $e^{\pm jk_{\rm n} z}$ of these waves, and $z$ is the
axis orthogonal to the interfaces. It is known that the T-matrix
of a serial connection of several devices described by their
T-matrices is simply a multiplication of the matrices in the order
determined by the connection.

The total transmission matrix from the
source  plane to the plane where the source field distribution
is ideally reconstructed must be the identity matrix
\e
T_{\rm total} = T_{\rm space\ after}\.T_{\rm device}\.T_{\rm space\ before} =
\left(\begin{array}{cc}
1 & 0\\
0 & 1\\
\end{array}\right) \l{Ttot} \f
for  every spatial harmonic of the
source field. Here, $T_{\rm space\ before}$ and $T_{\rm space\
after}$ represent air layers occupying the space between the source plane
and the device, and the space between the device and the image plane.
$T_{\rm device}$ is the transmission matrix of our device.
From this formula it is obvious that
a complete reconstruction of the field  distribution in the source plane at a distant image
plane must involve phase compensation for the propagating space
harmonics and ``amplification'' for the evanescent ones. In other
words, we need to synthesize a device that somehow inverts the action
of a free space layer.

Condition \r{Ttot} is a strict condition requiring not only
the device one-way transmission to be such that it reconstructs
the source field picture at the image plane, but also the matching
to be ideal and the device operation to be symmetric (reversible
in the optical sense).
We will consider
some less strict
conditions a bit later.
The matrices presented in \r{Ttot} can be written in explicit forms.
A space layer of thickness $d/2$ has the T-matrix
\e T_{\rm space} = \left(\begin{array}{cc}e^{-jk_{\rm n}d/2} & 0\\
0 & e^{+jk_{\rm n}d/2}\\\end{array}\right) \f
To compensate the
action of two such layers before and after the device, the device
T-matrix $T_{\rm device}$ has to be, obviously, the inverse of the
transmission matrix of these space layers:
\e T_{\rm device} =
\left(\begin{array}{cc}e^{+jk_{\rm n}d} & 0\\ 0 & e^{-jk_{\rm
n}d}\\\end{array}\right) \l{Tideal} \f
There are at least two solutions known for this
idealistic case by now: an ideal Veselago slab lens and a system of two
conjugating planes discussed above\footnote{Perhaps, it is not easy to see directly why \r{Tideal} holds
for a pair of conjugating planes, since there is only the normal component
of the wave vector in \r{Tideal} which is kept untouched by~\r{BC}.
But considering conjugations at both planes together and taking the inner
space partial wave propagators $e^{\pm jk_{\rm n}d}$ into
account, equation \r{Tideal} can be easily obtained.}.
Are other solutions possible? Let us consider a device that is a combination of two
``field transformers" (e.g., this sheets of certain
electromagnetic properties) separated
by a layer of free space. In other words, we want to study if we
can replace conjugating planes by some other layers, hopefully more
easily realizable. This device is modeled by the transmission matrix
\e T_{\rm device} =
\left(\begin{array}{cc}
a & b\\
c & d\\
\end{array}\right)\.
\left(\begin{array}{cc}
e^{-jk_{\rm n}d} & 0\\
0 & e^{+jk_{\rm n}d}\\
\end{array}\right)\.
\left(\begin{array}{cc}
e & f\\
g & h\\
\end{array}\right) \l{great} \f
Here, the first and the last matrices with yet unknown components
describe the two layers forming the device.
It is easy to show that if
$a=d=0$, $e=h=0$ and $bg=cf=1$ then the total device T-matrix
takes form  \r{Tideal},
i.e., the necessary matrix of an ideal lens.

The next question is whether there is
a way to realize a layer with the transmission matrix
of the form
\e T_{\rm trans} = \left(\begin{array}{cc}
0 & b\\
c & 0\\
\end{array}\right) \l{Ttrans} \f
A person with experience in
microwave engineering would  probably say that such T-matrix
can never be achieved in a physical
device, because it corresponds to a scattering matrix (well-known
S-matrix) having all components being infinite. However, a more
careful investigation of this question will lead us to an important
result presented in the next section.

\section{The use of impedance grids}

Here, we will show
that under certain restrictions  devices that ``amplify'' evanescent fields
can be designed using only passive elements, if we do not demand that the
same device reconstructs the propagating part of the spectrum.
This is possible because the two main phenomena on the two
interfaces in a perfect lens (negative refraction, necessary to focus
propagating modes, and surface polariton resonance, necessary to
reconstruct the evanescent spectrum) are fundamentally different, and
the required properties of such sheets are not necessarily combined in a
single design.

Let us consider a simple system: a lossless isotropic grid, e.g., a
conductive wire mesh. If the grid induced current is only electric current,
and there is no effective magnetic current induced in the grid
(e.g., when the grid structure is completely planar), then the grid
reflection coefficient $R$ and transmission coefficient $T$  at
the grid plane are connected as
\e T = 1 + R \f
provided that they
are defined through the electric field tangential components. The
corresponding T-matrix of such a grid is
\e T_{\rm g} =
\left(\begin{array}{cc}
\ds {1+2R\over 1+R} & \ds {R\over 1+R}\\[3mm]
\ds -{R\over 1+R} & \ds {1\over 1+R}\\
\end{array}\right)
\f

It is possible to make grids supporting propagation of surface
modes (also known as {\it slow waves} in radio engineering). For wire meshes, for
example, this phenomenon was investigated in \cite{Kontorovich}.
If the tangential component of the wave vector of an incident wave
coincides with the propagation factor of a surface mode, the
surface mode resonance appears. Obviously, the incident wave
should be evanescent in this case to match with the propagation constant of the
surface mode.
At a surface mode resonance $R\rightarrow\infty$ (for evanescent
modes $R$ is not bounded by $|R| \le 1$). Then, the grid T-matrix
takes the form
\e T_{\rm g} = \left(\begin{array}{cc}
2 & 1\\
-1 & 0\\
\end{array}\right) \l{Tg} \f
It is almost of the necessary form
\r{Ttrans}. Remembering that  conditions~\r{Ttot} and
\r{Tideal} are too strict in many cases, let us calculate the
total device matrix using~\r{Tg} directly. We obtain
\e T_{\rm
device} = T_{\rm g}\.T_{\rm space}\.T_{\rm g} =
\left(\begin{array}{cc}
4e^{-jk_{\rm n}d} - e^{jk_{\rm n}d} & 2e^{-jk_{\rm n}d}\\
-2e^{-jk_{\rm n}d} & -e^{-jk_{\rm n}d}\\
\end{array}\right) \f
which seems to be far from the desired, but
let us check what is the device S-matrix. After some algebra
involving the known relations connecting the elements of T and S matrices:
\e S = \left(\begin{array}{cc}
-t_{21}/t_{22} & 1/t_{22}\\
t_{11} - t_{12}t_{21}/t_{22} & t_{12}/t_{22}\\
\end{array}\right)
  \f
we get
\e
S_{\rm device}=
\left(\begin{array}{cc}
-2 & -e^{+jk_{\rm n}d}\\
-e^{+jk_{\rm n}d} & -2\\
\end{array}\right) \l{Sd} \f
Notice pluses in the exponents for
$s_{12}$, $s_{21}$. They mean that the  device can ``amplify''  resonating
evanescent modes. This is yet another indication of the fact
that Pendry's amplification \cite{Pendry} means the resonance growth in a surface mode
resonance.
The device based on simple grids considered above is not a lens,
though. It cannot focus propagating modes, as equation \r{Sd}
holds only at surface mode resonances ($R = \infty$). Another
``imperfectness'' of the found realization is that there is no
ideal matching in this case. The incident resonant evanescent mode
reflects from the device with the coefficient $-2$.

The matrix components of \r{Sd} can lead to various possible
design conditions, less strict than \r{Ttot} or \r{Tideal}. For
example, a direct analogy gives the following form of the device
S-matrix
\e S_{\rm device}\propto \left(\begin{array}{cc}
r & e^{+jk_{\rm n}d}\\
e^{+jk_{\rm n}d} & r\\
\end{array}\right)
\f
for a symmetric reciprocal device. The corresponding T-matrix for this case reads
\e
T_{\rm device}\propto
\left(\begin{array}{cc}
e^{jk_{\rm n}d} - r^2 e^{-jk_{\rm n}d} & r e^{-jk_{\rm n}d}\\
-r e^{-jk_{\rm n}d} & e^{-jk_{\rm n}d}\\
\end{array}\right)
\f
There are various other possibilities involving asymmetric and nonreciprocal devices.

\section{Conclusions}

In this paper, general requirements and possibilities for
realization of planar lenses being able to reconstruct near-field
distributions of a source with sub-wavelength resolution (so-called
perfect lenses) have been considered. Based on the observation
that the phenomena at the two boundary surfaces of a slab lens are
more critical for a lens operation than the propagation phenomena
inside the lens material, we have arrived to the following main
conclusions.

One does not necessarily need a composite medium possessing
negative $\E$ and $\M$ or another kind of backward-wave medium to
produce a perfect lens. An analogous device can be constructed
using various other possibilities. So far, two general
possibilities have been considered. One is based on usage of two
parallel artificially made surfaces or sheets imposing boundary
conditions of  form \r{BC} on fields in free space. The realizability of
such a device is not forbidden (at least for single-frequency, or
steady-state, operations) by physical laws and we are looking
forward for using nonlinear (or active) materials to achieve this
purpose. Operating as a planar lens, the device is able to focus
propagating modes of a source providing in the same time an
``amplification'' of the evanescent modes, i.e. it is able to work
as a perfect sub-wavelength resolution imaging device.

The second possibility lies in a wide class of planar structures
supporting slow waves. The surface mode (polariton) resonance
occurs in such structures when the incident field wave vector
tangential component coincides with the propagation factor of a
slow wave. It happens if the incident wave is an evanescent wave
in free space. We have shown that using surface mode resonances in
lossless grids placed in air it is possible to achieve
``amplification'' of evanescent modes, like in the case of
Veselago's slab lens. This phenomenon can be used not only
for making optical or electromagnetic imaging devices more precise, but
also for applications where the surface mode
resonance allows to detect small field irregularities in a near
field of an object.

\end{document}